\documentclass[twocolumn,aps,prl,preprintnumbers,showpacs,nofootinbib]{revtex4}
\usepackage[utf8]{inputenc}
\usepackage[english]{babel}
\usepackage{amsmath}
\usepackage{amsfonts}
\usepackage{amssymb}
\usepackage{epsfig}
\usepackage{graphics,psfrag,rotating}
\usepackage{graphicx}
\usepackage{dcolumn}
\usepackage{bm}
\usepackage{epstopdf}
\usepackage{color}
\usepackage[usenames,dvipsnames,svgnames]{xcolor}
\usepackage[colorlinks=true,
            linkcolor=red,
            urlcolor=gray,
            citecolor=blue]{hyperref}
  \usepackage{hyperref}

\newcommand{\lsim}   {\mathrel{\mathop{\kern 0pt \rlap
{\raise.2ex\hbox{$<$}}}
 \lower.9ex\hbox{\kern-.190em $\sim$}}}
\newcommand{\gsim}   {\mathrel{\mathop{\kern 0pt \rlap
{\raise.2ex\hbox{$>$}}}
\lower.9ex\hbox{\kern-.190em $\sim$}}}

\def\3nab{\tilde{\nabla}}

\def\tl{\tilde}
\def\hsp5{\hspace{5mm}}

\def\case#1/#2{\textstyle\frac{#1}{#2}}

\def\ber {\begin{eqnarray}}
\def\eer {\end{eqnarray}}
\def\bea {\begin{eqnarray}}
\def\eea {\end{eqnarray}}

\def\bc {\begin{center}}
\def\ec {\end{center}}
\def\case#1/#2{\frac{#1}{#2}}

\newcommand{\bw}{\begin{widetext}}
\newcommand{\ew}{\end{widetext}}
\newcommand{\nn}{\nonumber\\}

\newcommand{\del}{\Delta}
\newcommand{\be}{\begin{equation}}
\newcommand{\bse}{\begin{subequation}}
\newcommand{\ese}{\end{subequation}}
\newcommand{\ee}{\end{equation}}
\newcommand{\eei}{\end{eqnarray}\indent\indent}
\newcommand{\ba}{\begin{array}}
\newcommand{\ea}{\end{array}}
\newcommand{\bal}{\begin{eqnarray}}
\newcommand{\eal}{\end{eqnarray}}

\newcommand{\hs}{\,-\,}

\newcommand{\car}{{\cal R}}

\def\case#1/#2{\textstyle\frac{#1}{#2} }

\newcommand{\bver}{\begin{verbatim}}
\newcommand{\ever}{\end{verbatim}}

\def\cqg{{\em Class. Quantum Grav.\/} }
\def\grg{{\em Gen. Rel. Grav.\/} }
\def\prd{{\em Phys. Rev.\/} {\bf D}}

\def\aph{{\em Ann. Phys. (NY)\/} }
\def\plb{{\em Phys. Lett.\/} {\bf B}}

\begin{document}
\title{Large Scale Structure Constraints for a Class of $f(R)$ Theories of Gravity}
\author{Amare Abebe$^{1,2}$, \'Alvaro de la Cruz-Dombriz$^{1,2,3}$, Peter K. S. Dunsby$^{1,2,4}$}
\affiliation{$^{1}$ Astrophysics, Cosmology and Gravity Centre (ACGC), University of Cape Town, Rondebosch, 7701, South Africa}
\address{$^{2}$  Department of Mathematics and Applied Mathematics, University of Cape Town, 7701 Rondebosch, Cape Town, South Africa}
\address{$^{3}$  Instituto de Ciencias del Espacio (ICE/CSIC) and Institut d'Estudis Espacials de Catalunya (IEEC), Campus UAB, Facultat de Ci\`{e}ncies, Torre C5-Par-2a, 08193 Bellaterra (Barcelona) Spain}
\address{$^{4}$  South African Astronomical Observatory,  Observatory 7925, Cape Town, South Africa.}
\date{\today}
\begin{abstract}
Over the last few years much attention has been given to the study of modified gravity theories in order to find a more natural explanation for the late time acceleration of the Universe. Nevertheless, a comparison of the matter power spectrum predictions made by these theories with available data has not yet been subjected to a detailed analysis. In the context of $f(R)$ theories of gravity we study the predicted power spectra using both a dynamical systems approach for the background and solving for the matter perturbations without using the quasi-static approximation, comparing the theoretical results with several SDSS Data. 
The importance of studying the first order perturbed equations by assuming the correct background evolution and the relevance of the initial conditions are 
also stressed.  We determine the statistical significance in relation to the observational data and demonstrate their conflict with existing observations.
\end{abstract}
 \maketitle
\section{Introduction}
Despite more than a century of careful scrutiny, General Relativity (GR) remains the best fundamental theory of physics for describing
gravitational interactions. When applied to cosmology,  assuming that the large-scale geometry of the Universe is given by the Friedmann-La\^{i}matre-Robertson-Walker (FLRW) metric and adopting a fluid description for the matter content consisting of baryons, Cold Dark Matter (CDM) and radiation, GR gives rise to a set of field equations which can be solved exactly to give the simplest expanding Universe model - the so called  Friedmann model, governing the dynamics of the cosmological scale-factor $a(t)$.  This model has been remarkably successful, giving for example the correct abundances of the lightest elements and explaining the origin of the Cosmic Microwave Background Radiation (CMBR). In the last two decades, however, advances in observational cosmology appear to suggest that if one wishes to retain the FLRW metric, the  Universe must have undergone two periods of accelerated expansion. The first period is needed to explain the flatness problem and the near-scale invariant spectrum of temperature fluctuations observed in the CMBR,  while the second period explains the dimming of distant type Ia supernovae relative to Einstein-de Sitter Universe model.  In order to explain these periods of acceleration, the strong energy condition ($\mu+3p\geq 0$) needs to be violated. In the case of inflation, this is achieved by introducing a dynamical scalar field, while the present day acceleration is most easily explained with the introduction of a positive cosmological constant. The resulting description of the Universe has become known as the $\Lambda$CDM or {\em Concordance Model} \cite{concordance}. Although this beautifully simple phenomenological model appears to fit all currently available observations (Supernovae Ia \cite{sneIa}, CMBR anisotropies \cite{cmbr}, Large Scale Structure formation \cite{lss}, baryon acoustic oscillations \cite{bo} and weak lensing \cite{wl}), it is affected by significant fine-tuning problems related to the vacuum energy scale and this has led in recent years to efforts to explore alternatives to this description of the Universe.

The leading alternative to the $\Lambda$CDM model is based on modifications of the standard Einstein-Hilbert action. This is due to the fact that these changes naturally admit a phase of late time accelerated expansion (an early universe inflationary phase is also possible \cite{star80}). In this way Dark Energy can be thought of as having a geometrical origin, rather than being due to the vacuum energy or additional scalar fields which are added by hand to the energy momentum tensor (see \cite{DEfR} and references therein for an extensive presentation of the state of the art of this programme of investigation).

One of the simplest extensions of GR is based on gravitational actions which are non-linear in the Ricci curvature $R$ and$/$or contain terms involving combinations of derivatives of $R$ \cite{DEfR, ccct-ijmpd,otha,cct-mnras,kerner, teyssandier,magnanoff}. An important feature of these theories is that the field equations can be written in a way which makes it easy to compare with GR. This is done by moving all the higher-order corrections to the curvature onto the RHS of the field equations and defining an "effective" source term, known as the {\em curvature fluid}. Once this has been done the strong energy condition can be easily violated and this gives rise to a curvature fluid-driven period of late-time acceleration. Unfortunately this comes at the cost of having to study a considerably more complex set of field equations, making it difficult to obtain both exact and numerical solutions which can be compared with observations. Many studies of the expansion history of $f(R)$ gravity and other modified gravity theories have been performed using a range of strategies for numerically integrating the cosmological equations and these studies have highlighted, among other things, how sensitive the results are to initial conditions, the presence of rip singularities and oscillations in the deceleration and snap parameters, the existence and stability of Einstein Static models and bounce solutions \cite{background-collection}.

Theses difficulties can be reduced somewhat by using the theory of dynamical systems \cite{DS}, which, with careful choice of dynamical variables, provides a relatively simple method for obtaining exact solutions (via the equilibrium points of the system) and a description of the global dynamics of these models for a given $f(R)$ theory \cite{Dyn}.

Another useful (but more limited) approach is to assume that the expansion history of the Universe is known exactly, and to invert the field equations to deduce what class of $f(R)$ theories give rise to this particular cosmological evolution {\bf \cite{DobadoPRD}}. This has been done recently for exact power-law solutions for the scale factor, corresponding to phases of cosmic evolution when the energy density is dominated by a perfect fluid. It was found that such expansion histories only exist for modifications of the type $R^n$ \cite{julien}. It was also shown in \cite{reconstruction1} that the only $f(R)$ theory of gravity that admits an exact $\Lambda $CDM expansion history is standard General Relativity with a positive cosmological constant and if one wants to obtain this behaviour of the scale factor for more general functions of $R$, additional degrees of freedom need to be added to the matter sector.
A more extensive analysis of {\em reconstruction methods} has been carried out in \cite{odintsov} to obtain theories which give an approximate description of deceleration-acceleration transitions in cosmology and also in  \cite{reconstruction2} where the reconstruction method was based on standard cosmic parameters instead of specifying the time evolution of the scale factor.

Because it is possible to find background expansion histories which are consistent with the standard $\Lambda$CDM model, it is necessary to investigate the growth of structure in order to break this degeneracy. This requires extending the standard theory of cosmological perturbations for GR to $f(R)$ gravity. This has been done using both the metric based approach \cite{dombriz, Starobinsky, Silvestri,Qstatic_Bean} originally developed for GR by Bardeen \cite{bardeen} and the $1+3$ covariant approach first introduced by Ellis and Bruni \cite{covariant}. For example, in \cite{perts}, evolution equations were obtained for scalar and tensor perturbations for $f(R)$ gravity and applied to the spatially flat, matter dominated solution of $R^n$ gravity given by $a(t)=a_0t^{\frac{2n}{3(1+w)}}$ (which is the saddle point $G$ of the corresponding dynamical system for these theories). 

The results obtained demonstrate that the evolution of scalar perturbations is determined by a fourth order differential equation, so that the evolution of density fluctuations contains, in general, four modes rather than the standard two obtained in GR. This results in more complex perturbation dynamics than what is found in GR. It was also found that the perturbations depend on the scale for any value of the equation of state parameter for standard matter (while in GR the evolution of the dust perturbations is not scale dependent) and that there is a scale-dependent feature in the matter power spectrum \cite{PRL}.  Furthermore,  the growth of large density fluctuations can occur also in backgrounds which have a negative deceleration parameter. These surprising results are very different to what one finds in GR and could be used to constrain some $f(R)$ theories using the integrated Sachs-Wolfe (ISW) effect \cite{Zhang} and the matter power spectrum \cite{comment}.

These features can be interpreted by comparing the system of fourth order equations, which produced them, with the corresponding equations for two interacting fluids in GR \cite{DBE},  because they have the same structure, i.e., there are friction and source terms due to the interaction of the two effective fluids. On very large and on very small scales, the system of equations become independent of $k$, so that the evolution of the perturbations does not change as a function of scale and the power spectrum is consequently scale invariant. On intermediate scales,  the curvature fluid acts as a relativistic component, whose pressure is responsible for the oscillations and the dissipation of the small scale perturbations in the same way photons operate in a baryon-photon system. This suggests that the variables describing the fluctuations in the curvature fluid can be interpreted as representing the modes associated with the additional scalar degree of freedom typical of $f(R)$-gravity.  In this way, the spectrum can be explained physically as a consequence of the interaction between the additional scalar modes and those resulting from standard matter. This leads to a large drop of power when the parameter $n$ is varied by a small amount.

The results described above were obtained assuming the simple background model $G$ and can therefore only provide hints about what features to expect in the matter power spectrum when realistic expansion histories are considered. In order to test the robustness of the results presented in \cite{perts}, the complete expansion history of the background FLRW universe for $R^n$ gravity, which resembles the $\Lambda$CDM model needs to be found. This is achieved by integrating the dynamical systems equations so that an orbit representing a cosmic history passes close to the matter dominated point $G$ and eventually tends towards the late-time attractor $C$. In \cite{Dyn} it was shown that if $1.36<n<1.5$, $G$ and $C$ respectively represent a decelerated matter dominated solution and late-time power-law acceleration. The equations for $\Delta_m$ and ${\cal R}$ can then be solved for this background to obtain a matter power spectrum which can be directly compared with the one found for $\Lambda$CDM.
 
Other attempts have tried to encapsulate the effects of the extra $f(R)$-terms in the involved equations by parameterising some relevant functions.  Inspired by the 
behaviour $f(R)$ models in the quasi-static limit,  Bertschinger and Zukin \cite{Bertschinger_parametric}  originally parameterised the two gravitational potentials in terms of a time and scale-dependent Newton's constant  and the so-called gravitational slip. 
Under these assumptions numerical codes which compute the growth of cosmological perturbations have been implemented. For instance several parametrizations 
using MGCAMB  \cite{Parametric} and more recently CLASSgal  \cite{CLASSgal_Durrer} are available.
  
 
This paper is organised as follows: In Section \ref{Section2} we give the basic equations governing the background evolution of FLRW models and briefly present the key variables needed to study cosmological perturbations in $f(R)$ gravity using the covariant approach.  In Section \ref{Section3} we discuss the cosmology of $R^n$ gravity and describe their main features by recasting the cosmological equations as an autonomous system of first order equations. We then integrate these equations for different values of the parameter $n$ using initial conditions in the radiation dominated epoch which have Hubble and deceleration parameters equal to their $\Lambda$CDM values. For such initial conditions, we compare the cosmological background evolution with baryon acoustic oscillations (BAO) data in Section  \ref{Section4}. We find that it is impossible to have cosmic histories that simultaneously provide Hubble and deceleration parameters close to the $\Lambda$CDM values today. We also find that the BAO analysis corroborates the inviability of these models at the cosmological evolution level. In Section \ref{Section5} we determine the matter power spectra for the expansion histories given in \ref{Section3} and compare them with what is obtained by integrating the perturbation equations for the matter dominated solution $G$. We find that although the broad large and small scale features of the power spectrum are largely the same as in \cite{PRL}, the scale-dependent features are no longer present when the complete background expansion history is considered.  We then use the observed matter power spectrum based on both luminous red galaxies (LRG) \cite{SDSS} 
and the DR9 CMASS galaxy sample observed by the Sloan Digital Sky Survey (SDSS)-III \cite{Ashley} 
to directly constrain this class of $f(R)$ theories of gravity. We find that 
the models considered give power spectra in the SDSS-III wavenumber interval, which are in good 
agreement with the available data for the recent DR9 CMASS sample. 

Finally in Section \ref{Section6} we discuss the results and give an outline of future work to be done.

Unless otherwise specified, natural units ($\hbar=c=k_{B}=8\pi G=1$)
will be used throughout this paper, Latin indices run from 0 to 3.
The symbol $\nabla$ represents the usual covariant derivative and we use the
$-,+,+,+$ signature and the Riemann tensor is defined by
\begin{eqnarray}
R^{a}{}_{bcd}=W^a{}_{bd,c}-W^a{}_{bc,d}+ W^e{}_{bd}W^a{}_{ce}-
W^f{}_{bc}W^a{}_{df}\;,
\end{eqnarray}
where the $W^a{}_{bd}$ are the Christoffel symbols (i.e., symmetric in
the lower indices), defined by
\begin{equation}
W^a{}_{bd}=\frac{1}{2}g^{ae}
\left(g_{be,d}+g_{ed,b}-g_{bd,e}\right)\;.
\end{equation}
The Ricci tensor is obtained by contracting the {\em first} and the
{\em third} indices
\begin{equation}\label{Ricci}
R_{ab}=g^{cd}R_{cadb}\;.
\end{equation}
Finally the action for $f(R)$-gravity can be written in these units as:
\begin{equation}\label{lagr f(R)}
\mathcal{A}=\int {\rm d}^4 x \sqrt{-g}\left[\frac12 f(R)+{\cal L}_{m}\right]\;,
\end{equation}
where $R$ is the Ricci scalar, $f$ is general differentiable 
(at least $C^2$) function of the Ricci scalar and $\mathcal{L}_m$ corresponds to 
the matter Lagrangian. 
\section{The Cosmological Equations for $f(R)$  Gravity} \label{Section2}
\subsection{The Background FLRW equations}
In a FLRW universe, the  non-trivial field equations lead to the following equations governing the expansion history of the Universe:
\begin{eqnarray}\label{raych}
&&3\dot{H}+3H^2=-\frac{1}{2f'}\left[\mu_m+3p_m+f-f'R+3H f'' \dot{R}\right.\nonumber\\ 
&&~~~~~~~~\left.+3f'''\dot{R}^2+3f''\ddot{R}\right]\label{ray}\;,\\
&&3H^2= \frac{1}{f'}\left[\mu_m+\frac{Rf'-f}{2}-3H f'' \dot{R}\right]\;, \label{fried}
\end{eqnarray}
i.e., the {\em Raychaudhuri} and {\em Friedmann} equations \cite{RF}. Here $H$ is the Hubble parameter, which defines the scale factor $a(t)$ via the standard relation $H=\dot{a}/{a}$, the Ricci scalar is 
\begin{equation}
R=6\dot{H}+12H^2 \label{Riccci}
\end{equation}
and  $f',f''$ and $f'''$  abbreviate $\partial^n f/{(\partial R)^n}$ for $n=1..3$ respectively. The {\em energy conservation equation} for standard matter
\begin{equation}\label{cons:perfect}
\dot{\mu}_m=-3H\mu_m\left(1+w\right)
\end{equation}
closes the system, where $w$ is the barotropic equation of state.

Note that the Raychaudhuri equation can be obtained from the Friedmann equation, the energy conservation equation and the definition of the Ricci scalar. Hence, any solution of the Friedmann equation automatically solves the Raychaudhuri equation. 
\subsection{Density perturbations in $f(R)$ gravity}
Density (scalar) perturbations may be extracted from any first order tensor $A_{ab}$ orthogonal to the 4\hs velocity $u^a$ by using a  {\em local} decomposition \cite{covariant}, so that repeated application of the operator $D_{a}\equiv h^{b}{}_{a}\nabla_b$ on $A_{ab}$ extracts the scalar part of the perturbation variables. In this way we can define the following scalar quantities
\begin{eqnarray}
&&\Delta_{m}=\frac{a^2}{\mu_{m}}D^2\mu_{m}\,,~~~
Z=3a^2D^2H\,,~~~ C=S^{4}D^2\tilde{R}\,,\nonumber\\
&&{\cal R}=a^2D^2 R\,,~~~\Re=a^2D^2\dot{R}\;,
\end{eqnarray}
where $h_{ab}$ is the projection tensor into the rest-spaces orthogonal to $u^a$.  $\Delta^m_a$,  $Z$ respectively represent the fluctuations in the matter energy density $\mu_m$ and expansion $\Theta$, and ${\mathcal R}$,  $\Re$ determine the fluctuations in the Ricci scalar $R$ and its momentum $\dot{R}$. This set of variables completely characterises the evolution of density perturbations.  Then, using eigenfunctions of the  spatial Laplace-Beltrami operator defined in \cite{covariant}: $D^{2}Q = -\frac{k^{2}}{a^{2}}Q$,
where $k=2\pi a/\lambda$ is the wavenumber and $\dot{Q}=0$,  we can expand every first order quantity in the above equations, so for example in the case of $\Delta_m$ we have
\begin{equation}\label{eq:developmentdelta}
\Delta_m(t,\mathbf{x})=\sum \Delta^{(k)}_m(t)\;Q^{(k)}(\mathbf{x})\;,
\end{equation}
where $\sum$ stands for both a summation over a discrete index or an integration over a continuous one. In this way, it is straightforward, although lengthy, to derive a pair of second order equations describing the $k^{th}$ mode for density perturbations in $f(R)$ gravity. They are:
\bw
\begin{eqnarray}
&&\ddot{\del}^{k}_{m}-\left[\left(3w-2\right)H +\frac{\dot{R}f''}{f'} \right]\dot{\del}^{k}_{m}+\left[ w\frac{k^{2}}{a^{2}}+(w-1)\frac{\mu_{m}}{f'}-w\frac{f}{f'}\right]\del^{k}_{m}\nonumber\\
&&~=\frac{1+w}{2}\left[-1-\frac{2k^{2}}{a^{2}}\frac{f''}{f'}+(f-2\mu_{m}+6\dot{R}H f'')\frac{f''}{f'^{2}} -6\dot{R}H\frac{f'''}{f'}\right]\car-\frac{3(1+w)}{f'}H f''\dot{\car}^{k},\nonumber\\
&&\ddot{\car}^{k}+\left(2\dot{R}\frac{f'''}{f''}+3H\right)\dot{\car}^{k}+\left[\frac{k^{2}}{a^{2}}+\ddot{R}\frac{f'''}{f''}+\dot{R}^{2}\frac{f^{(iv)}}{f''}+3H\dot{R}\frac{f'''}{f''}+\frac{f'}{3f''}-\frac{R}{3}\right]\car^{k}\nn
&&~=-\left[\frac{1}{3}(3w-1)\frac{\mu_{m}}{f''}+\frac{w}{1+w}\left(2\ddot{R}+2\dot{R}^{2}\frac{f'''}{f''}+6\dot{R}H\right) \right] \del^{k}_{m}+\frac{1-w}{1+w}\dot{R}\dot{\del}^{k}_{m}\;.
\label{sys3}
\end{eqnarray}
\ew
Already on super-Hubble scales, $k/aH\ll 1$, a number of important features can be found which allow one to differentiate from what is obtained in GR \cite{perts}. Firstly, it is clear that the evolution of density perturbations is determined by a {\it fourth order} differential equation rather than a second order one. This implies that the evolution of the density fluctuations contains, in general, four modes rather that two and can give rise to a more complex evolution than the one of GR.  Secondly, the  perturbations are found to depend on the scale for any equation of state for standard matter (while in GR the evolution of the CDM perturbations are scale-invariant). This means that even for dust, the evolution of super-horizon and sub-horizon  perturbations are different. Thirdly, it is found that the  growth of large density fluctuations can occur also in backgrounds in which the expansion rate is increasing in time. This is in striking contrast with what one finds in GR and would lead to a time-varying gravitational potential, putting tight constraints on the ISW for these models.

Let us now turn to the case of a general wave mode $k$.  One of the most instructive ways of understanding the details of the evolution of density perturbations for a general $k$ is to compute the matter transfer function $T(k)$, defined by the relation \cite{Coles} $\langle\Delta_{m}({\mathbf k_1})\Delta_{m}({\mathbf k_2})\rangle=T(k_1) \Delta({\mathbf k_1}+{\mathbf k_2})$, where ${\mathbf k_i}$ are two wavevectors characterizing two Fourier components of the solutions of (\ref{sys3}) and $T({\mathbf k_1})=T(k_1)$ because of isotropy in the distribution of the perturbations. This quantity tells us how the fluctuations of matter depend on the wavenumber at a specific time and carries information about the amplitude of the perturbations (but not on their spatial structure). In GR, the transfer function on large scales is constant, while on small scales it is suppressed in comparison with the large scales (i.e., modes which entered the horizon during the radiation era) \cite{PaddyPert}.  In the case of pure dust in GR the transfer function is scale invariant.  Substituting the details of the background, the values of the parameter $n$, the barotropic factor $w$ and the wavenumber $k$ into (\ref{sys3}) one is able to obtain $T(k)$ numerically.

One can easily see from expressions \eqref{sys3}  that the matter power-spectrum in $f(R)$ gravity theories 
is further processed after equality and would differ from the standard $\Lambda$CDM power spectrum 
$P_{k}^{\Lambda\text{CDM}}$  when evaluated today. The latter is widely assumed to represent accurately the evolution of perturbations till radiation-matter equality, since  before that the effects of any modification to the usual Concordance Model needs to be negligible in order to  preserve the cosmological standard model predictions in the radiations-dominated epoch such as the primordial light elements abundances during Big Bang Nucleosynthesis.

Therefore, these two power spectra, when evaluated today, would be related linearly by a transfer function $T(k)$ given by
\begin{eqnarray}
P_{k}^{f(R)}\,=\, T(k) P_{k}^{\Lambda\text{CDM}}|_{eq}\;,
\label{transfer}
\end{eqnarray}
where $T(k)^{}\propto|\del_{m}^{k}|^{2}_{today}$ and $\del_{m}^{k}$ is obtained from the system of equations (\ref{sys3}).
 
On linear scales, $P_{k}^{f(R)}$ will in general depend on both the $f(R)$-model and the scale $k$, therefore differing from the $\Lambda$CDM model, where it is scale-invariant.

In the GR limit: $f(R)=R$, (\ref{sys3}) reduces to the standard equations for the evolution density 
perturbations in GR:
\begin{eqnarray}
&&\ddot{\del}^{k}_{m}-\left(3w-2\right)H\dot{\del}^{k}_{m}\nonumber\\
&&+\left[w\frac{k^{2}}{a^{2}}+\left(\frac{-1+2w-3w^2}{2}\right)\mu_{m} \right] \del^{k}_{m}=0,
\end{eqnarray}
\begin{eqnarray}
\car^{k}=(1-3w)\mu_{m}\del^{k}_{m},
\end{eqnarray}
and one can easily see that the linear evolution of CDM density perturbations for sub-Hubble ($k \gg aH$) scales in $\Lambda$CDM is given by the well-known result:
\be
\del_{m}^{''k}+{\mathcal{H}}\del_{m}^{'k} - \frac{1}{2}a^2 \mu_{m} \del^{k}_{m}\,=\,0\;,
\label{delta_0_SubHubble}
\ee
where ${\cal H}=a'/a$ and prime (only for this equation) denotes derivative with respect to conformal time. 

Notice that according to (\ref{delta_0_SubHubble}) the evolution of the Fourier modes does not depend upon $k$. This means that for $\Lambda$CDM models on sub-Hubble scales, once the density contrast starts to grow after matter-radiation equality, evolution only changes the overall normalisation of the matter power-spectrum $P(k)$,  but not its shape. 
\section{Determining the Expansion History for $R^n$-gravity} \label{Section3}
To proceed, we need to fix our theory of gravity. The simplest and most widely studied form of $f(R)$ gravitational theories is  $f(R)=\alpha H_0^2(R/H_0^2)^n$, where $\alpha=\alpha(n)$ is a non-dimensional coupling constant and $H_0$ is the $\Lambda$CDM value of the Hubble  parameter today. For this class of models the cosmological equations associated with a FLRW universe are particularly easy to analyse.

However, the aim of this investigation is to show that studies of different $f(R)$-gravity models that share a similar background expansion history with the $\Lambda\text{CDM}$ model can in principle provide useful constraints on the viability of these models  via the power spectra of matter density perturbations they produce and with the help of BAOs as standard rulers of known geometrical information.

The first step in the implementation of the Dynamical Systems (DS) approach for determining the expansion history for $R^n$ gravity is the definition of the key DS variables. Following \cite{Dyn}, we introduce the dimensionless variables:
\begin{eqnarray}
&&x =\dfrac{\dot{R} (n-1)}{H R},\;\;y=\dfrac{R(1-n)}{6nH^2},\nn
&&\Omega_d=\dfrac{\mu_{d}}{3n \alpha H^2 R^{n-1}},\;\;\Omega_r=\dfrac{\mu_{r}}{3n \alpha H^2 R^{n-1}}\,.
\label{DV}
\end{eqnarray}
In terms of these variables, the Friedmann equation (\ref{fried}) takes the form
\begin{eqnarray}\label{const}
1+x+y-\Omega_d-\Omega_r=0\;.
\end{eqnarray}
An autonomous system of ordinary differential equations, which are equivalent to cosmological equations (\ref{raych}-\ref{cons:perfect}) can be obtained by differentiating (\ref{DV}) with respect to redshift $z$. Here we give the equations for dust and radiation, while those for a general barotropic equation of state $w$ are presented in \cite{Dyn}:
\begin{eqnarray}
-(z+1)\frac{\text{d}x}{\text{d}z}&=&-x-x^2+\dfrac{(4-2n+nx)y}{n-1}+\Omega_d\,,\nonumber\\
-(z+1)\frac{\text{d}y}{\text{d}z}&=&4y+\dfrac{(x+2ny)y}{n-1}\,,\nonumber\\
-(z+1)\frac{\text{d}\Omega_d}{\text{d}z}&=&\left(1-x+\dfrac{2ny}{n-1}\right) \Omega_d\nn
-(z+1)\frac{\text{d}\Omega_r}{\text{d}z}&=&\left(-x+\dfrac{2ny}{n-1}\right) \Omega_r\;.
\label{DS2}
\end{eqnarray} 
The dimensionality of the resultant system (\ref{DS2}) can be reduced further using the Friedmann constraint (\ref{const}). 
The evolution of the Hubble parameter can then be determined by writing (\ref{Riccci}) in terms of the DS variables:
\be
(1+z)\frac{\text{d}h}{\text{d}z}=\frac{h(2+ny)}{n-1}\;,
\ee
where $h=H/H_0$. Furthermore the deceleration parameter can be determined directly from $y$:
\be
q=\frac{ny}{(n-1)}+1\;.
\ee
In \cite{Dyn} it was shown that these equations admit a number of fixed points of which two are particularly interesting. The points, labeled $G$ and $C$ in \cite{Dyn} correspond to two cosmologically interesting exact solutions: $G$ represents a matter dominated saddle point, which in the case of dust has $a=a_0t^{2n/3}$ and $C$ is the late-time attractor with $a=a_0t^{\frac{(1-n)(2n-1)}{n-2}}$. In \cite{Dyn} it was also shown that $C$ and $G$ respectively represent decelerated and accelerated phases of the Universe with positive energy density if $n$ lies in the range $1.36<n<1.5$.

With this in mind let us integrate (\ref{DS2}) by fixing the initial conditions for the DS variables (\ref{DV}) to be identical to their $\Lambda$CDM values  in the radiation dominated era (at a redshift $z=6000$) and determine the expansion history for $R^n$ models with eight different values for the exponent $n>1$ between $n= 1.1$ and $1.4$ in order to allow the possibility of late-time acceleration. In this way we can determine for which values of $n$ we obtain present day values for $q(z)$ and $H(z)$ consistent with the $\Lambda$CDM model. 
It is clear from the results in Table \ref{icss} 
that it is not possible for $R^n$ gravity to admit FLRW cosmic histories that simultaneously have present-day values of the Hubble and deceleration parameters close to their $\Lambda$CDM values today if initial conditions are chosen in order that the expansion history is close to the $\Lambda$CDM model at early times.


\begin{center}
\begin{table}[ht!]
\begin{tabular}{||c|c|c|c|c|c|c|c|c|c||}
\hline
\hline
$n$ & $1.1$ & $1.2$ & $1.27$ & $1.29$ & $1.3$& $1.31$& $1.33$&  $1.4$\\
\hline
$h_{0}$& $0.65$ & $0.75$& $0.94 $&$0.99$ &$1.44$ & $2.43$ & $ 7.34$ &$159.67$\\
\hline
$q_0$ &$0.39$ &$0.20$&$ 0.10$  &$0.25$ &$ 0.36$& $0.35$& $ 0.22$ &$-0.17$  \\
\hline
\hline
\end{tabular}
\caption{
Present-day values of the Hubble $h_0\equiv H({\rm today})/H_0$ and deceleration ($q_0$) parameters for the $R^n$ models under consideration. $H_0$ corresponds to the $\Lambda$CDM Hubble parameter value today. Only $n=1.4$ provides acceleration at the present time, whereas  $n=1.29$ gives the closest value for $h_0$ to $\Lambda$CDM. With regard to the $\chi^2$ analysis for BAO to be studied in the Section \ref{Section4} $n=1.29$ provided the best value ($\chi^2_{\rm BAO}=16.11$) but well above the one provided by $\Lambda$CDM ($\chi^2_{\rm BAO}=4.51$). 
The remaining values of exponent $n$ give $\chi^{2}_{\rm BAO}$ values showing incorrect fits to BAO data.
}
\label{icss}
\end{table}
\end{center}

\section{BAO constraints} \label{Section4}
As standard rulers, BAO constraints provide an ideal arena in the analysis of cosmic expansion history. This is mainly because these oscillations correspond to a preferred length scale in the early universe that can be predicted from CMB measurements \cite{wigglez}. Some relevant quantities for these analyses are the comoving distance from an observer to some redshift $z$ which is given by
\be
r(z)=\frac{1}{H_0}\int^{z}_{0}\frac{{\rm d}z}{h(z)}\;,
\ee 
the scaled distance to recombination,
 the comoving sound horizon at recombination and the dilation scale respectively are given by \cite{lazkoz}
\ber
&&R\, =\, H_0\sqrt{\Omega_{0d}}\; r(z_{\rm CMB})\;,\\
&&r_s(z_{\rm CMB})\,=\,\frac{1}{H_0}\int_\infty^{z_{\rm CMB}}\frac{c_s(z)}{h(z)}\, {\rm d}z\;,\\
&& D_V
(z_{\rm BAO})\,=\,\left[\left(\int_0^{z_{\rm BAO}} \frac{{\rm d}z}{H(z)}\right)^2
\frac{z_{\rm BAO}}{H(z_{\rm BAO})}\right]^{1/3} 
\eer
where $c_s(z)=\left[3\left(1+\frac{\bar
{R}_b}{1+z} \right)\right]^{-1/2}$ is the sound speed of the photon-baryon relativistic plasma  with photon-baryon density ration 
\be 
\bar {R_b}\,=\,
\frac{3}{4}\frac{\Omega_b \tl h^2}{\Omega_{\gamma}\tl h^2}= 3.15\times 10^{4}\,\Omega_b \tl h^2 \left(\frac{T_{\rm CMB}}{2.7\, {\rm K}}\right)^{-4}.
\ee
Here $\tl h$ is the Hubble uncertainty  parameter defined by $H_0=100\tl h$ and we have used the Planck result of $\tl h=0.6711$ \cite{planckpar} for this analysis, as well as $z_{CMB}=1021.44$. 

Following the methods presented in \cite{lazkoz} 
 we study, for the different $n$ values considered, the BAO data likelihood 
 corresponding to recent measurements \cite{baoplanck} of the 6dF Galaxy Survey at $z=0.1$ \cite{beutler11}, the SDSS DR7  at $z=0.2, ~0.35$ \cite{percival10, padman12},  the WiggleZ at $z=0.44, 0.60, 0.73$ \cite{ wigglez}. Thus we define 
 %
%
\begin{eqnarray}
\bf{X_{BAO}} &=& \left(\begin{array}{c}
\frac{r_s(z_{\rm CMB})}{{D_V(0.106)}}  - 0.336 \\
\frac{r_s(z_{\rm CMB})}{{D_V(0.2)} }- 0.1905\\
\frac{r_s(z_{\rm CMB})}{{D_V(0.35)}}  -0.1097 \\
\frac{r_s(z_{\rm CMB})}{{D_V(0.44)} }- 0.0916\\
\frac{r_s(z_{\rm CMB})}{{D_V(0.6)}}  - 0.0726 \\
\frac{r_s(z_{\rm CMB})}{{D_V(0.73)} }- 0.0592\end{array} \right)\;.
\end{eqnarray}
%
to calculate the $\chi^2$ from the BAO as 
\be
\chi^2_{{\rm BAO}}=\bf{X_{BAO}}^{T}\,\,{\bf C_{BAO}}^{-1}\,\,\bf{X_{BAO}}\;. 
\ee
where ${\bf C_{BAO}}^{-1}$ corresponds to the inverse covariance matrix as given in \cite{wigglez}. 
The results found for the models under study showed that the $\chi^2_{\rm BAO}$ analysis proves that 
the cosmological evolution as provided by the models under study cannot achieve the goodness of $\Lambda$CDM ($\chi^2_{\rm BAO}=4.51$) and that only for $n=1.29$ 
($\chi^2_{\rm BAO}=16.11$) the  fit to BAO data can be considered of the same order of magnitude, though much bigger, than $\Lambda$CDM. In fact, for the model interval $n=[1.1,\,1.4]$ the $\chi^2_{\rm BAO}$ minimum lies at $n=1.29$ being the $\chi^2_{\rm BAO}$ value strongly dependent on the exponent $n$ so that for other values of $n$ the obtained $\chi^2$'s rapidly departed from this minimum.

\section{The Matter Power Spectrum and SDSS constraints} \label{Section5}
Let us now turn to the matter power spectrum. 
Taking the dominant component to be dust, the system (\ref{sys3}) can be written in terms of the dynamical variables:
\begin{widetext}
\begin{eqnarray}
&&\frac{(-1+n)^{2}(1+z)}{2ny}\hat{\car}^{k'}-\frac{3h^{2}\left[(-1+n)(1+(-2+n)\Omega_{d})+(-2+n)y\right]+\hat{k}^2(-1+n)^{2}(1+z)^{2}}{6h^{2} n y}\hat{\car}^k
\nonumber\\
&&+h^{2}(1+z)^{2}\Delta_{m}^{k''}+h^{2}\frac{\left[(-1+n)\Omega_d+y\right](1+z)}{n-1}\Delta_{m}^{k'}-3h^{2}\Omega_{d}\Delta_{m}^{k}=0,\nonumber\\
\label{sys3_bis}
&&\hat{\car}^{k''}-\frac{\left[4-4\Omega_d+4y+n\left(-2+2\Omega_d-3y\right)\right]}{(-1+n)(1+z)}\hat{\car}^{k'}  
+\left\{\frac{\hat{k}^2}{h^2}+\frac{(-2+n)\left[-\Omega_{d}^2-(-1+y)^{2}+\Omega_{d}(1+n+2y)\right]}{(-1+n)^2(1+z)^2}\right\} \hat{\car}^{k}\nonumber\\
&&+\frac{6h^{2}ny\left(1-\Omega_{d}+y\right)}{(-1+n)^{2}(1+z)}\Delta_{m}^{k'}+\frac{6h^2n\Omega_{d}y}{(-1+n)^2(1+z)^2}\Delta_{m}^{k}=0\;,
\end{eqnarray}
\end{widetext}
where prime denotes derivative with respect to redshift and the dimensionless quantities
 $\hat{\car}^k=\car^k/H_0^2$ and $\hat{k}=k/H_0$ have been introduced. Note that equations (\ref{sys3_bis}) are valid only for $n\neq1$.

SDSS correlation data either from LRG or from DR9 have been used to test the predictions from the $\Lambda$CDM power spectrum obtained from linear perturbation theory 
to high accuracy. For instance, $\chi^2\approx 11.2$, degrees of freedom ($d.o.f.=14$) for LRG \cite{SDSS} and $\chi^2 = 61.1$, ($d.o.f.=59$) for DR9 \cite{Anderson}. In what follows, we will do the same for this class of $f(R)$ theories of gravity.  

To do this we first determine the cosmological background evolution as described in the previous section and then use these results to solve the system of equations  (\ref{sys3_bis}) in order to obtain the density contrast today. Then, by applying expression (\ref{transfer}) to these results, one can obtain the fully processed power spectra $P_{k}^{f(R)}$ for the above models, which can be compared to the $\Lambda$CDM predictions for LRG and DR9 data.

Before proceeding, let us mention that three sets of different initial conditions were considered for the system (\ref{sys3_bis}) in order to determine how sensitive the final processed power spectrum is to changes in these values:
\begin{itemize}\label{icsss}
\item {\bf I}:  $\Delta_{m}^{k}|_{in}= \hat{\car}^{k}|_{in}=10^{-5}$,\;  $\Delta_{m}^{k'}|_{in}= \hat{\car}^{k'}|_{in}=10^{-5}$,
\item {\bf II}:  $\Delta_{m}^{k}|_{in}= \hat{\car}^{k}|_{in}=10^{-5}$,\;  $\Delta_{m}^{k'}|_{in}= \hat{\car}^{k'}|_{in}=10^{-8}$, 
\item {\bf III}: $\Delta_{m}^{k}|_{in}= \hat{\car}^{k}|_{in}=10^{-5}$,\;  $\Delta_{m}^{k'}|_{in}= \hat{\car}^{k'}|_{in}=0$, 
\end{itemize}
where the subscript $in$ refers to the initial redshift $z_{in}=2000$. The choice of sets {\bf I} and {\bf II} as initial conditions for the system
(\ref{sys3}) can be understood as providing scale-invariant initial conditions for the variables $\Delta_{m}^{k}$ and 
$\hat{\car}^{k}$ and their first derivatives are all taken to be small (but non-zero) at the initial redshift. 
On the other hand, in set {\bf III} we set the first derivatives of $\Delta_m^k$ and $\hat{\car}^{k}$ to zero at $z=z_0$. This choice has important consequences for the obtained spectra.

\begin{figure*}[htbp] 
	\centering
		\includegraphics[width=0.3250275\textwidth]{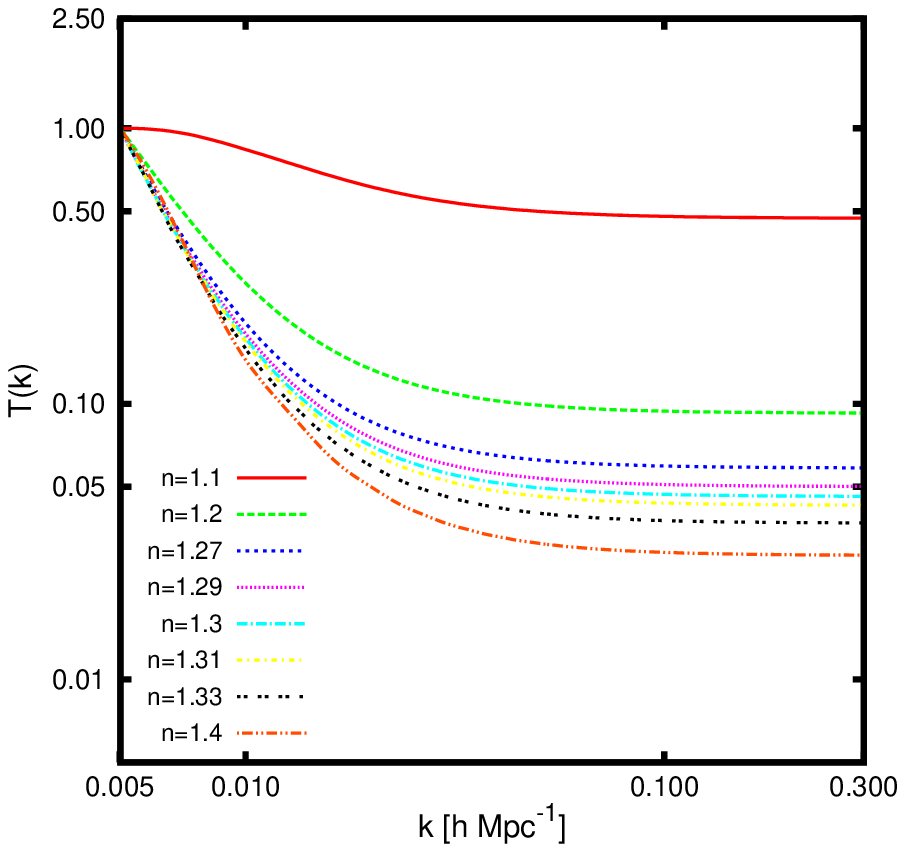}
		\includegraphics[width=0.3250275\textwidth]{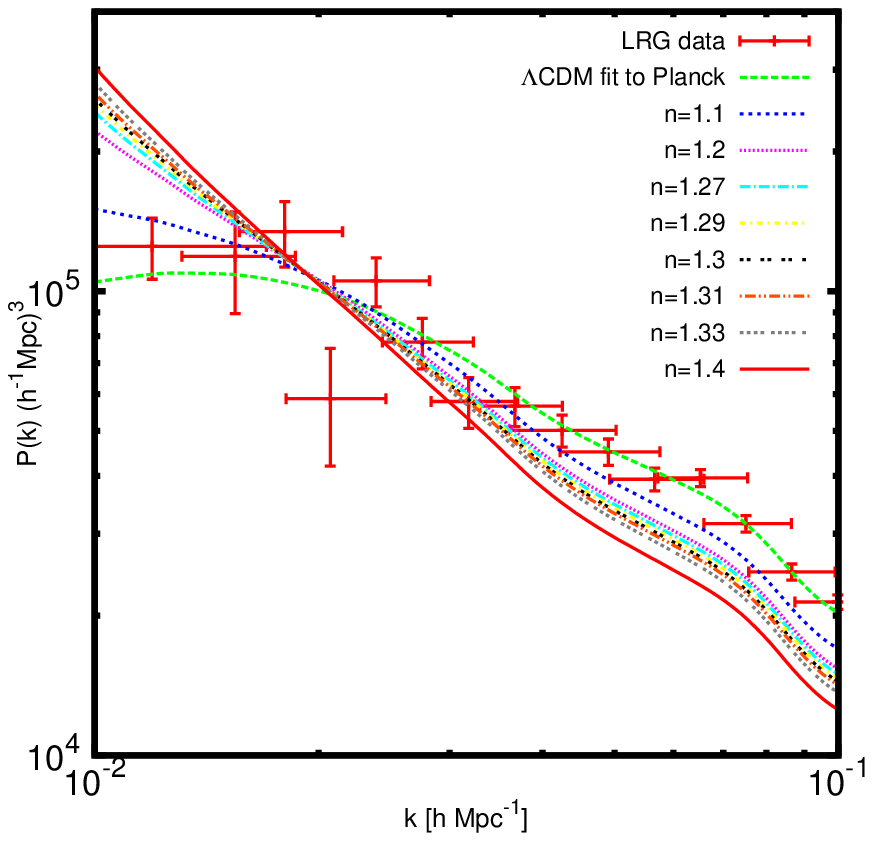}
		\includegraphics[width=0.3250275\textwidth]{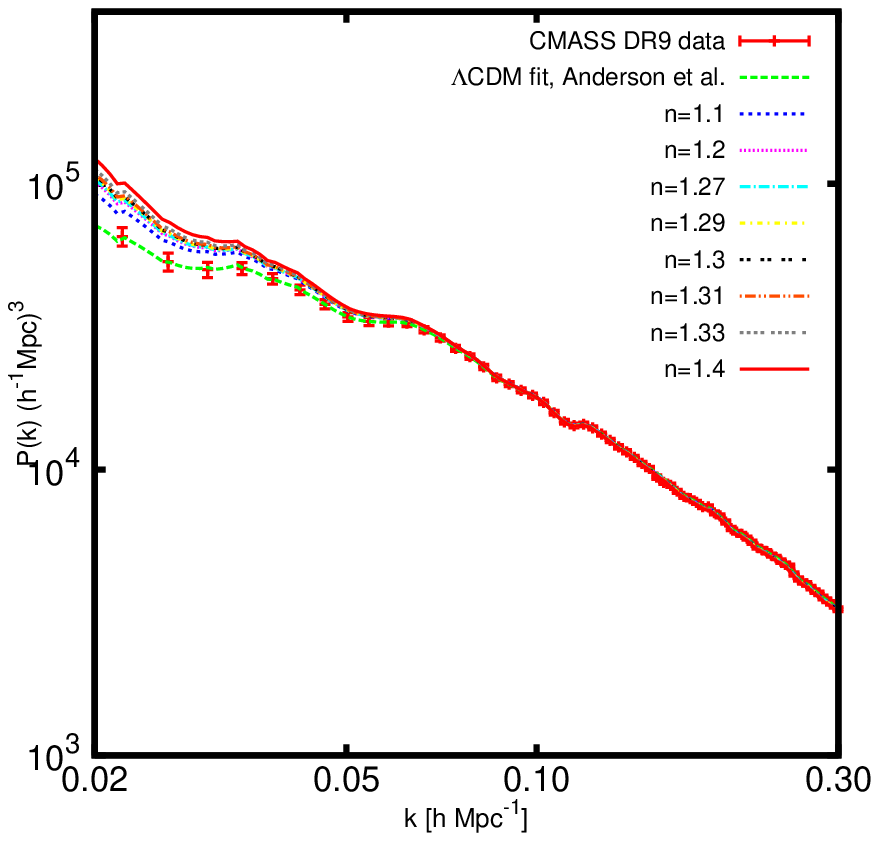}\\
		 \includegraphics[width=0.3250275\textwidth]{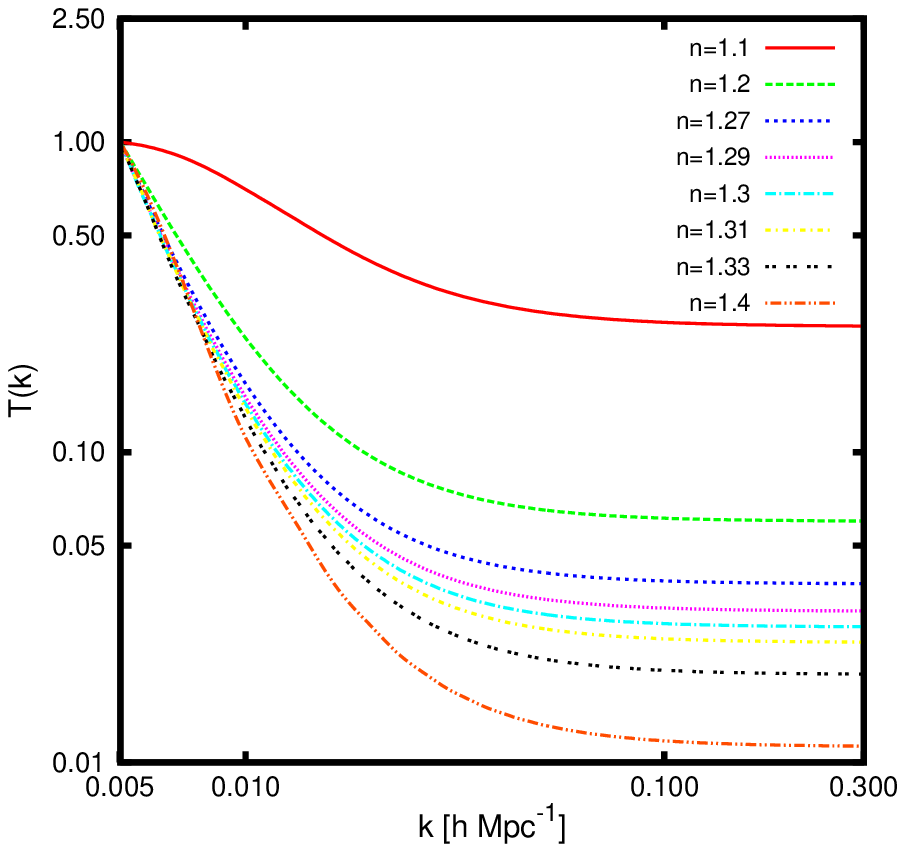}
		\includegraphics[width=0.3250275\textwidth]{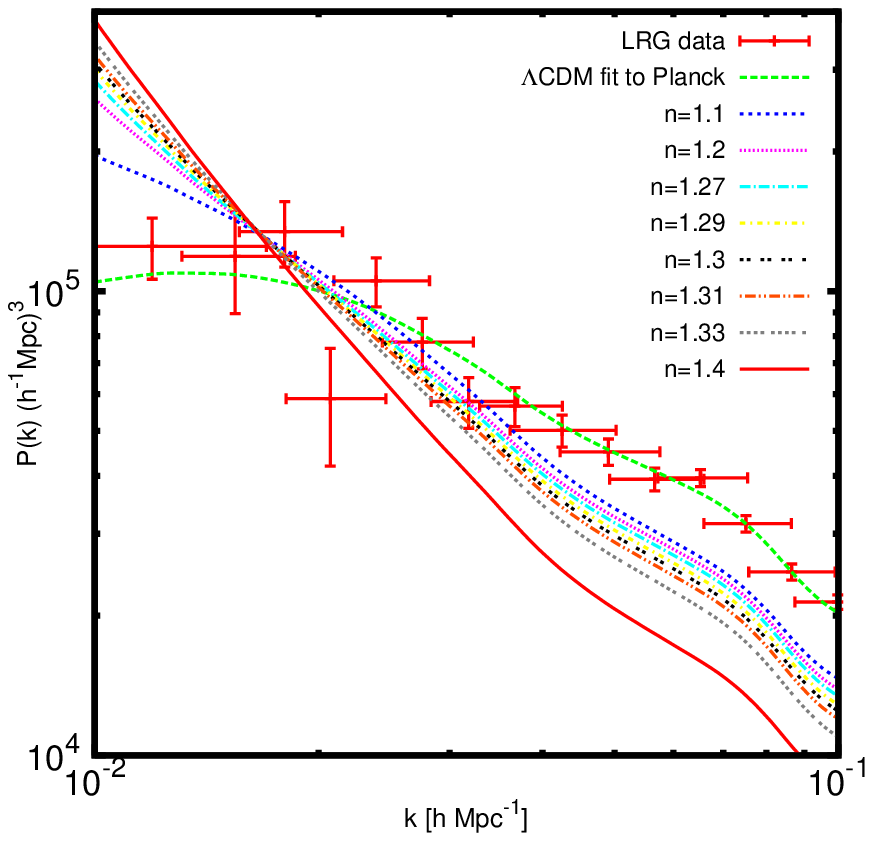}
		\includegraphics[width=0.3250275\textwidth]{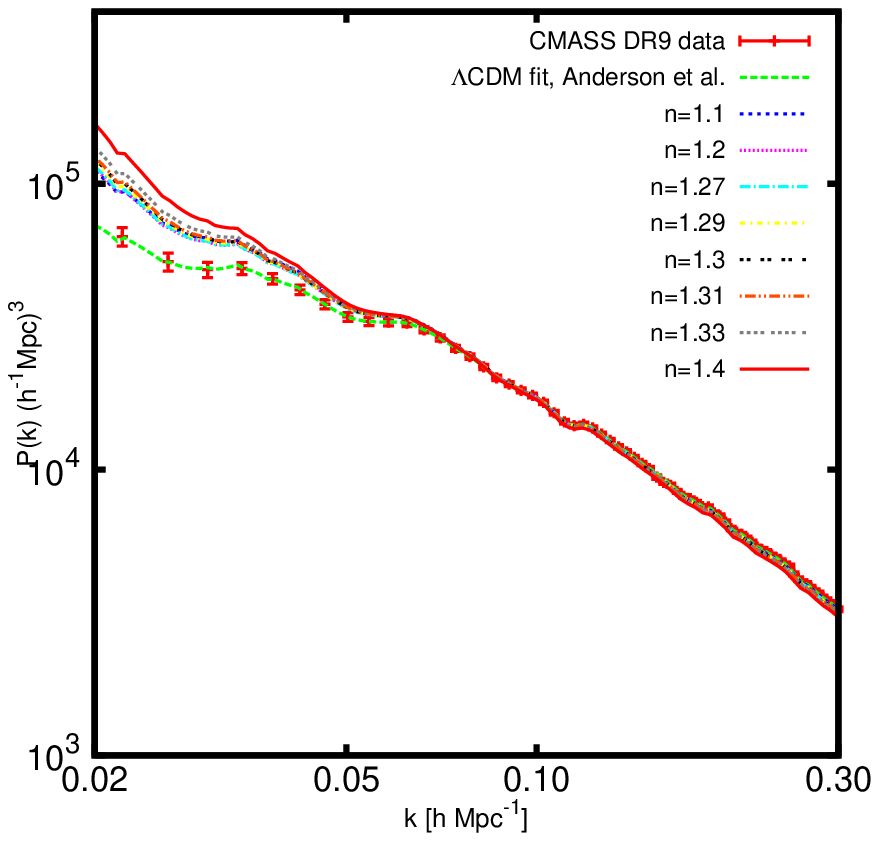}\\
          	\includegraphics[width=0.3250275\textwidth]{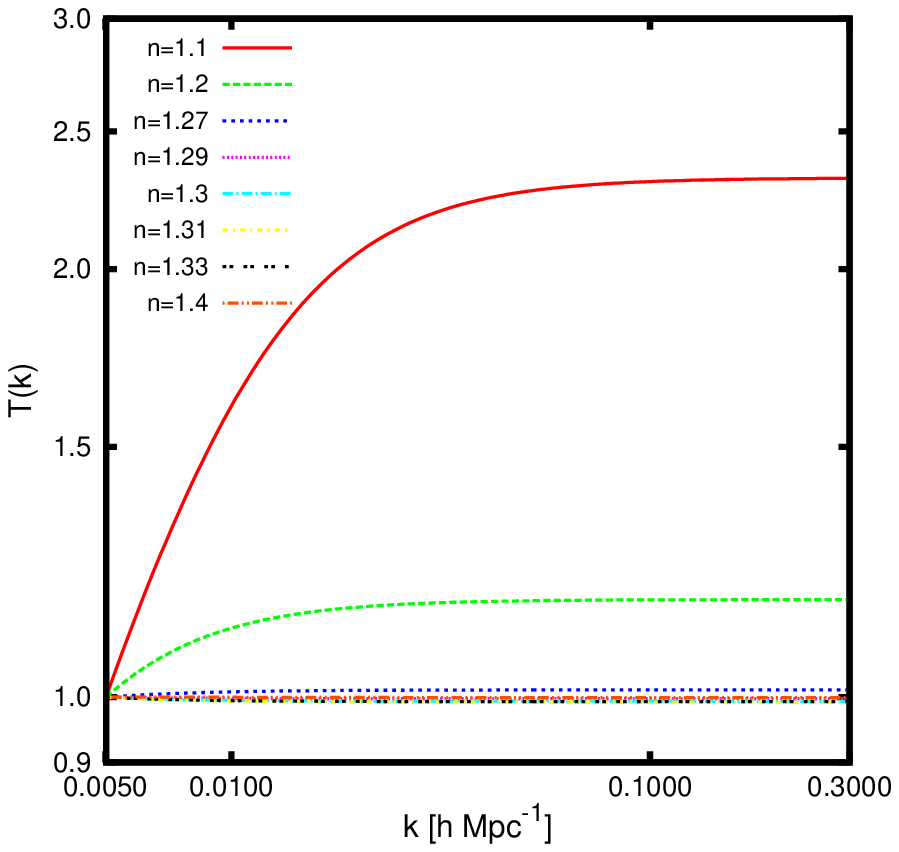}
		\includegraphics[width=0.3250275\textwidth]{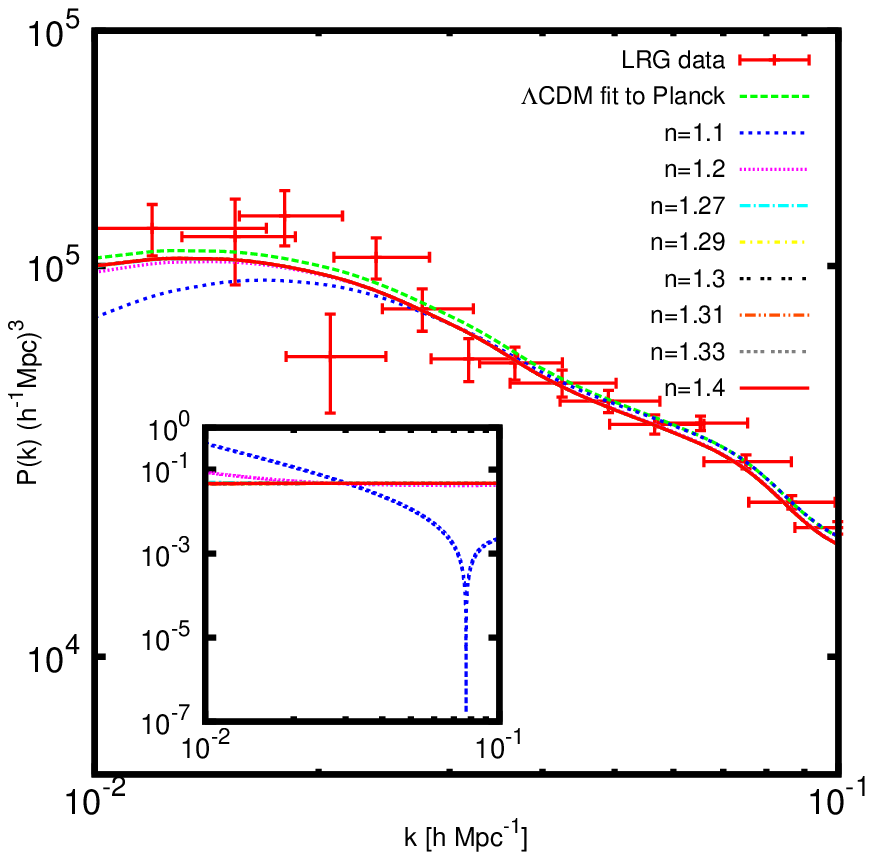}
		\includegraphics[width=0.3250275\textwidth]{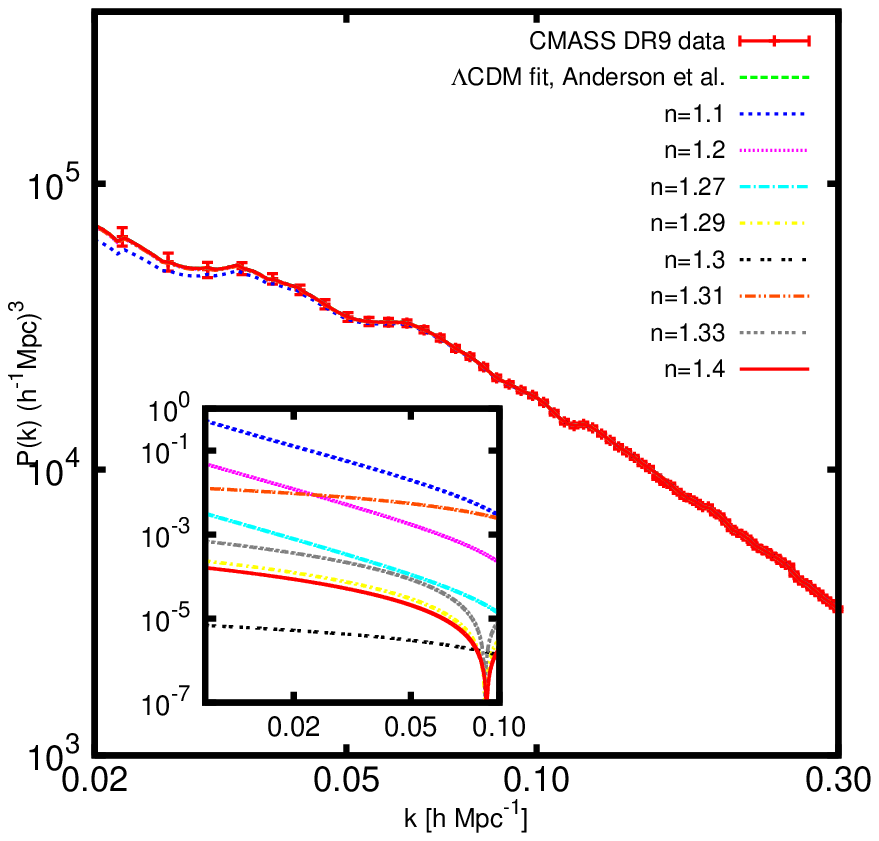}
 \caption{\footnotesize{ 
The left panels show the transfer function
$T(k)=|\Delta_{k}/\Delta^{\Lambda\text{CDM}}_{k}(z=2000)|^2$ evaluated today ($z=0$) for wavenumber $k$ (in $h\,\text{Mpc}^{-1}$ units) in the range $0.005$ to $0.3$ 
for the initial conditions sets {\bf I}, {\bf II} and {\bf III} as described in the bulk of this investigation. 
The transfer functions $T(k)$ on left panels have been normalised in such a way that the curves coincide on large scales. 
%
On the central and right panels we present the corresponding linear matter power-spectra $P(k)$
for $\Lambda\text{CDM}$ and $R^n$ models for 
$n= 1.1$, $1.2$, $1.27$, $1.29$, $1.3$, $1.31$, $1.33$ and $1.4$. 
Data correspond to SDSS 2006 \cite{SDSS} (central panel) and SDSS-III data \cite{Ashley} (right panel) respectively. 
All the power spectra were assumed to have an arbitrary overall normalisation at the scale $k=0.01\,h\,\text{Mpc}^{-1}$ (central panel) and 
$k=0.02\,h\,\text{Mpc}^{-1}$ (right panel) in order to find the best fit to the data. Conditions {\bf I} and {\bf II} lead to power spectra in complete disagreement with
the observed data. Conditions {\bf III}, due to the almost flatness of the spectra in the range covered by data present a good fit to the data.
On the bottom panels (central and right) we show in a window the 
relative discrepancy between the 
 $\Lambda\text{CDM}$ and the $R^n$ fits power-spectra for every studied exponent.
For SDSS 2006 data, the smallest discrepancy in scales $k>3\times 10^{-2}\,h\,\text{Mpc}^{-1}$ happens for $n=1.1$ whereas for smaller scales, all the remaining values of $n$ provide similar relative error around $5\times 10^{-2}$. 
For DR9 SDSS-III data, the smallest discrepancy (order $10^{-5}$) happens for $n=1.3$ in the whole scale-range despite some punctual values of $k$ where other exponents
may present smaller relative errors with respect to $\Lambda$CDM.
whereas for smaller scales, all the remaining values of $n$ provide similar relative error around $5\times 10^{-2}$.
 }}
 \label{Figure_SDSS}
\end{figure*}

For each value of $n$, we present in Fig. \ref{Figure_SDSS} both the transfer functions and the processed power spectra for all the initial conditions sets just mentioned. 
We can see that for these models the transfer functions have a nearly-flat plateau on large scales \cite{PRL} regardless of the set of initial conditions. On intermediate scales however, the density contrast behaviour and its amplitudes today depend both upon the value of $n$ and the initial conditions. The left panels in Fig. \ref{Figure_SDSS} clearly illustrate this fact.

Then, by using expression (\ref{transfer}) for the transfer function, one can obtain the processed power spectra and compare these theoretical results with the LRG data. We find that initial conditions {\bf I} and {\bf II} are not able to provide a good fit to the data catalogues due to the fact that for the required scales 
the spectra are not flat but significantly change with the wavenumber $k$ (see the left panels in Fig. \ref{Figure_SDSS}). On the other hand, initial conditions {\bf III} give power spectra which are in good agreement with the data. This is due to the almost-flat transfer function in the data range. Note however that the initial amplitude was assumed to be a free parameter which was determined to achieve the best fit. 

For this set of initial conditions,  Tables \ref{Table_chi_Exact_BG} and \ref{Table_chi_Exact_BG_2013}
show the $\chi^2$ 
analyses for the eight studied $R^n$ models when their respective spectra evolutions are fitted to the SDSS data. SDSS 2006 data are assumed to be non-correlated whereas correlations for DR9 data are given in \cite{Ashley}. We also include the value for the confidence regions $\sigma$ with respect to $\Lambda$CDM as well as the overall amplitude suppression in the initial scales 
to get the best fits after a Least Square method analysis. 
For the SDSS 2006 data, it is clear that none of the $R^n$ models under consideration acquire the same goodness of fit as the  $\Lambda$CDM model as seen in Table  \ref{Table_chi_Exact_BG} in the $\sigma$ exclusion regions. However, 
for the DR9 SDSS-III data, one can see that some  $R^n$ models, such as $n=1.3$ and also $n=1.27$, $1.29$, $1.33$ and $1.4$ provide competitive fits $\Lambda$CDM model as seen in Table \ref{Table_chi_Exact_BG_2013} in the $\sigma$ exclusion regions. 

\begin{widetext}
\begin{center}
\begin{table}[t!]
\begin{center}
\begin{tabular}{||c|c|c|c|c|c|c|c|c||}
\hline
\hline
$n$ exponent & $1.1$ & $1.2$ & $1.27$ & $1.29$ & $1.3$ & $1.31$ & $1.33$ & $1.4$\\
\hline
$\chi^2$  & 15.1394 & 13.1839 & 13.0184 & 13.0093 & 13.0104 & 13.0098 & 13.0102 &  13.0128 \\
\hline
$\sigma$ exclusion & 1.9849 & 1.4086 & 1.3486 & 1.3452 & 1.3457 & 1.3454 & 1.3456 & 1.3465 \\
\hline
\hline
\% suppression   & 29.5 & 7.94 & 4.75 & 4.45 & 4.37 & 4.33 & 4.35 & 4.47\\
\hline
\hline
\end{tabular}
\end{center}
\caption{
Fits to the SDSS 2006 data for $R^n$ cosmology by using set of initial conditions {\bf III}: eight different values of exponent $n$ were investigated from $n= 1.1$ to  $1.4$. Values for $\chi^2$ and the confidence region $\sigma$ are presented in the second and third rows respectively. The data to be fitted by the theoretical spectra are taken from \cite{SDSS}. 
The fit provided by $\Lambda$CDM ($\chi^2=11.1996$) is not improved by any of  these parameter values. The final row gives the suppression in the overall initial amplitude required to get the best fits. For all the values, this suppression turns out to be smaller than $30 \%$ and is therefore in the experimental uncertainty interval for this quantity. One can see that the best fit corresponds to the value $n=1.29$ with a suppression of $4.45\%$ and good fits are also obtained for $n=1.3$, $1.31$ and $1.33$ with similar suppressions.}
\label{Table_chi_Exact_BG}
\end{table}
\end{center}
\end{widetext}
\begin{widetext}
\begin{center}
\begin{table}[t!]
\begin{center}
\begin{tabular}{||c|c|c|c|c|c|c|c|c||}
\hline
\hline
$n$ exponent & $1.1$ & $1.2$ & $1.27$ & $1.29$ & $1.3$ & $1.31$ & $1.33$ & $1.4$\\
\hline
$\chi^2$  & 4.5463 & 1.0507 & 1.0366 & 1.0357 & 1.0355 & 1.0458 & 1.0360 &  1.0357 \\
\hline 
$\sigma$ exclusion & 1.874 & 0.123 & 0.0316 & 0.012 & 0.002 & 0.101 & 0.020 & 0.001\\
\hline
\hline
\% suppression  & 13 & 1.5 & 0.1 & 0.01 & 0.001 & 1 & 0.04 & 0.009 \\
\hline
\hline
\end{tabular}
\end{center}
\caption{
Fits to the SDSS CMASS DR9 data for $R^n$ cosmology by using set of initial conditions {\bf III}: eight different values of exponent $n$ were investigated from $n= 1.1$ to  $1.4$. Values for $\chi^2$ and the confidence region $\sigma$ are presented in the second and third rows respectively. The data to be fitted by the theoretical spectra are taken from \cite{Ashley}. 
The fit provided by $\Lambda$CDM ($\chi^2= 61.1/59 \approx 1.03559$) is slightly improved by  the $n=1.3$ parameter value. The final row gives the suppression in the overall initial amplitude required to get the best fits. For all the values, this suppression turns out to be smaller than $15 \%$ and is therefore in the experimental uncertainty interval for this quantity. For the best fit $n=1.3$ the corresponding suppression is $10^{-3}\%$ and very good fits are also obtained for $n=1.27$, $1.29$, $1.33$ and  $1.4$ with similar suppressions.}
\label{Table_chi_Exact_BG_2013}
\end{table}
\end{center}
\end{widetext}
%
%
%
%

%
%
\begin{figure*} [htbp] 
	\centering
		\includegraphics[width=0.329\textwidth]{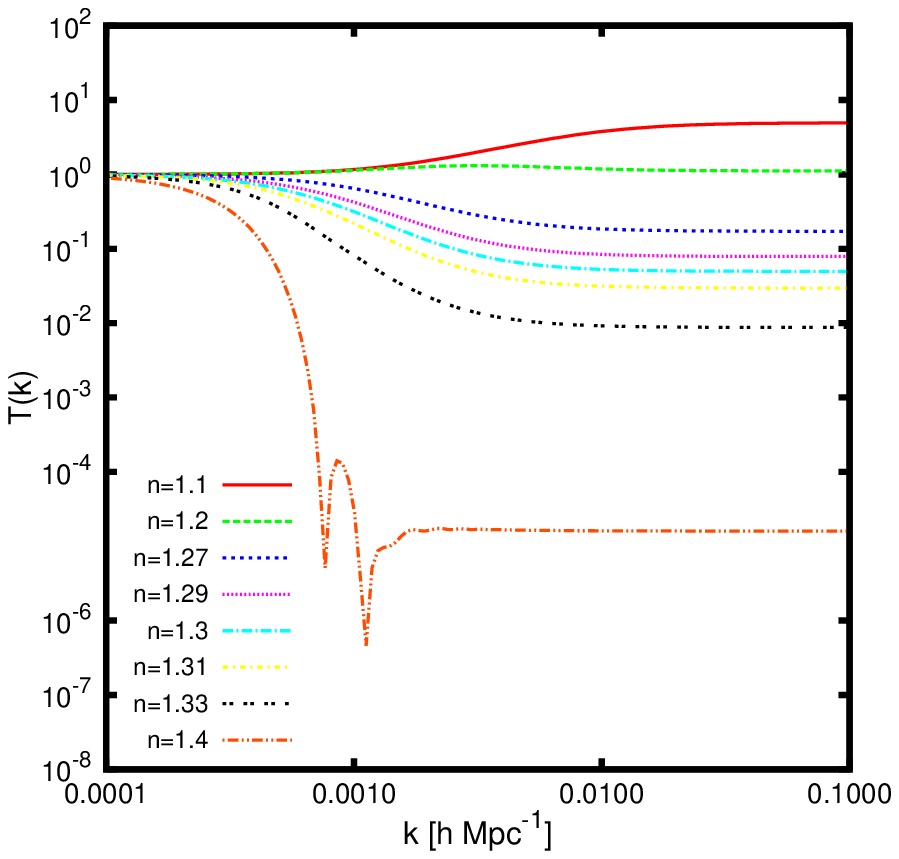}
		\includegraphics[width=0.329\textwidth]{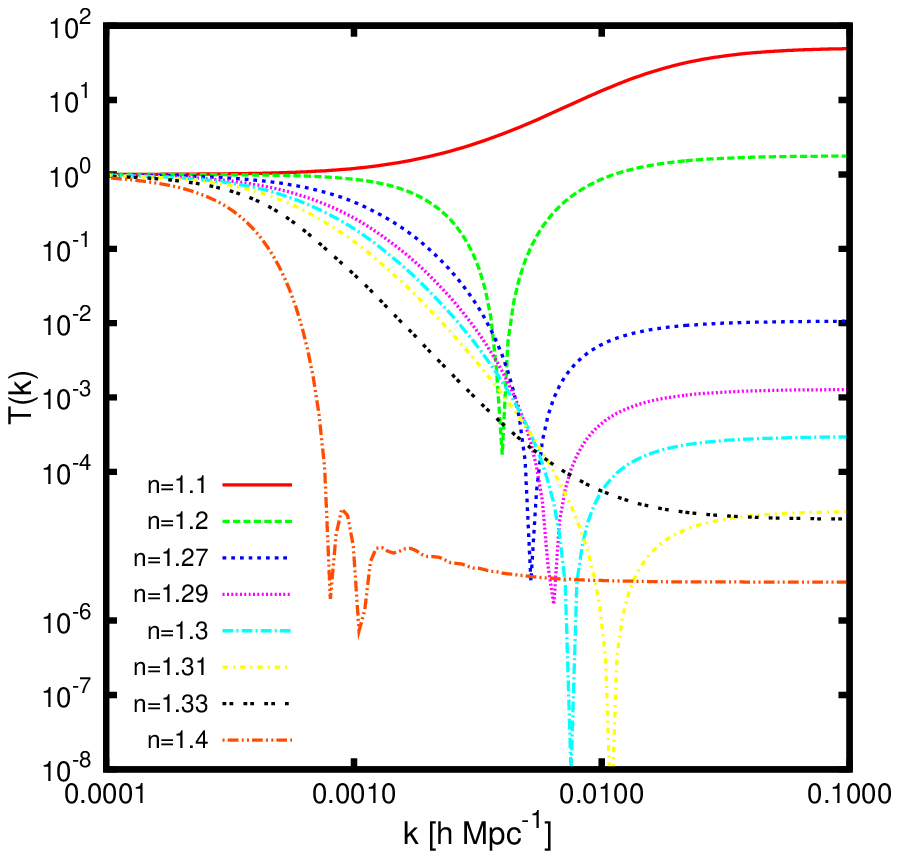}
		\includegraphics[width=0.329\textwidth]{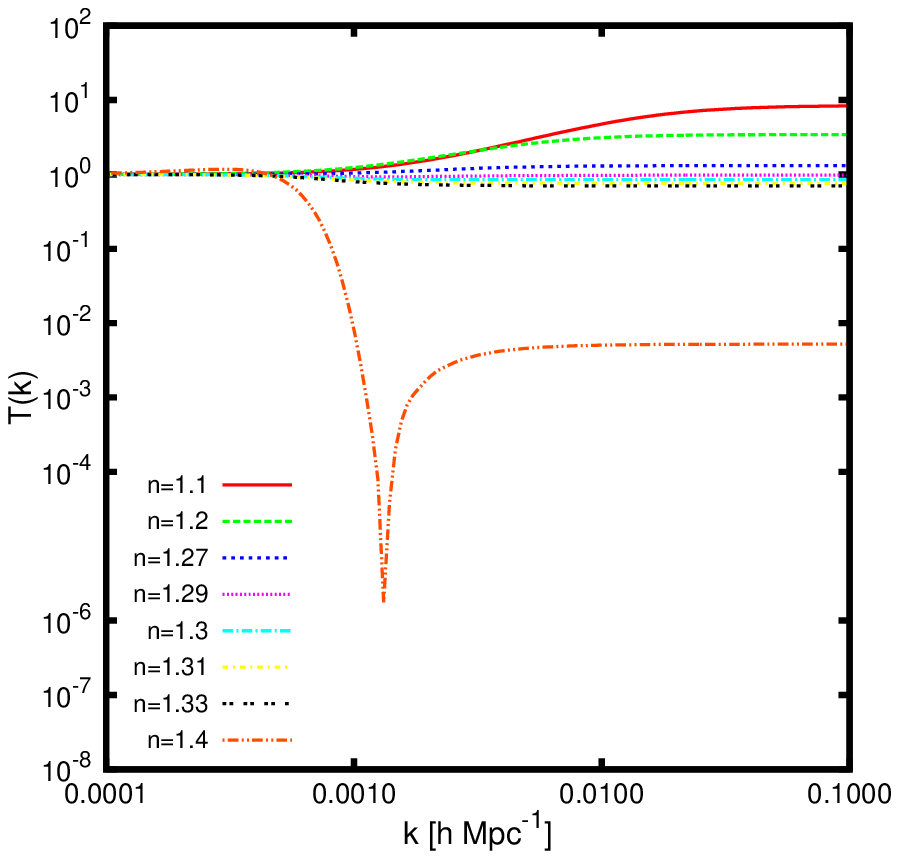}
		 \caption{\footnotesize{ 
Saddle-point analysis: scale dependence of the transfer function $T(k)=|\Delta_{k}/\Delta^{\Lambda\text{CDM}}_{k}(z=2000)|^2$ evaluated today ($z=0$) for wavenumber $k$ (in $h\,\text{Mpc}^{-1}$ units) in the range 
$0.0001$ to $0.1$  for the initial same conditions {\bf I}, {\bf II} and {\bf III} used in the rest of this investigation. 
The transfer functions have been normalized in such a way that the curves coincide on large scales. 
Both on very large and on very small scales, the $\Delta$ becomes $k$-independent, so that the evolution of
the perturbations does not change as a function of scale and the transfer function is consequently scale invariant.
On intermediate scales the curvature fluid causes the oscillations.  The result is a considerable
loss of power for a relatively small variation of the parameter $n$. As before the shape and amplitudes (i.e. the increasing or decreasing behaviour with $k$) of the transfer functions depend on the initial conditions {\bf I} (left), {\bf II} (center) and {\bf III} (right). For example, when  $n=1.27$ we have decreasing behaviour for {\bf I} and {\bf II} but increasing behaviour for {\bf III}.
 }}
 \label{Fig_saddle_point}
\end{figure*}

For completeness and in order to emphasise the importance of using the complete expansion history for the background, we have also given $|\Delta_m^k|^2$ for models whose background evolution is given by the exact saddle point solution $G$ in the case of dust ($w=0$). We do this for the same parameter values $n$ and initial conditions {\bf I}, {\bf II} and {\bf III} as shown in Fig. \ref{Fig_saddle_point}. These results agree with previous investigations \cite{PRL} that showed how when this background scale factor is assumed the spectrum is composed of three
parts corresponding to three different evolution regimes for the perturbations. In this scenario, on intermediate scales the interaction between the two fluids (dust and curvature) is maximised and the curvature fluid acts as a relativistic component whose pressure is responsible for the oscillations and the dissipation of the small scale perturbations in the same way in which the photons operate in a baryon-photon system \cite{PRL}. 
If we compare these results to the left panel in Fig. 1, we conclude that the scale-dependent features in Fig. 2 are washed out when the complete background expansion history is considered, however the main large and small scale features of the power spectrum found in \cite{PRL} are retained.

\section{Discussion and Future Work} \label{Section6}

In this paper we presented a complete analysis of the background and matter perturbations for one of the most widely studied modified gravity theories: $R^n$ gravity with $n\gtrsim 1$. Both the cosmological background evolution and linear perturbation equations were solved by combining the dynamical systems approach for the background and using the $1+3$ covariant approach to evolve the matter perturbations, without assuming any intermediate (quasi-static) approximation. 
 
We solved the background equations for different values of the parameter $n$ using initial conditions in the radiation dominated epoch, with Hubble and deceleration parameters equal to their $\Lambda$CDM values. For such initial conditions, we performed a baryon acoustic oscillations analysis. By using this tool we found that it is impossible to obtain fits as good as $\Lambda$CDM. We also proved the impossibility of having cosmic histories that simultaneously have present day values of these cosmological parameters close to their $\Lambda$CDM values today. In fact, of the ten models considered, only $n=1.4$ provided a negative deceleration parameter today, but gave a Hubble parameter completely incompatible with its observed value, while values around $n=1.29$ gave the closest value for the present-day Hubble parameter to $\Lambda$CDM but exhibits no late time acceleration. The value $n=1.29$ provided the best $\chi^2$ when its cosmological evolution is compared with BAO data but well above the 
$\Lambda$CDM one.

We then used the observed matter power spectrum based on both luminous red galaxies (2006) and the DR9 CMASS galaxy sample (2012) in the Sloan Digital Sky Survey to further constrain these models. For the studied exponents, we found that all the models gave rise to almost-flat transfer functions in the Sloan wavenumber interval provided very special initial conditions are chosen. In this case the best fit to the data for 2006 data was found for the value $n=1.29$ with a suppression of $4.45\%$ and good fits  and good fits were also obtained for $n=1.3$, $1.31$ and $1.33$. The exponent $n=1.4$ (the only one providing acceleration today) required a suppression slightly bigger ($4.47\%$). 
With regard to DR9 2012 data and partially thanks to the accuracy in this catalogue, most of the studied $R^n$ models provided good fits to the data being $n=1.3$ with a suppression of 
$10^{-3}\%$ the  best fit slightly improved by $\Lambda$CDM. Other exponents ($1.27$, $1.29$, $1.33$ and $1.4$) also provided good fits with slightly bigger suppressions. 
  
Regardless of the Large Structure Constraints none of the studied exponents were however able to fit the baryon acoustic oscillations data as well as the 
$\Lambda$CDM model and the obtained $\chi^2$ were much bigger than the best-fit model as provided by $\Lambda$CDM.
It is clear from this analysis that $R^n$ gravity does not successfully meet any of the cosmology requirements for it to be considered as a viable alternative to the standard model. This work does however illustrate in depth the utility of our approach and it should be possible to use these techniques with the most updated available data to constrain 
which $f(R)$ theories remain consistent with current data even if they are indistinguishable from the $\Lambda$CDM model either at the level of the FLRW background or cosmological perturbations.\\

{\bf Acknowledgments:} 
We would like to thank David Bacon for a comprehensive reading of the manuscript and for his useful comments. 
We are also indebted to Marc Manera and Ashley Ross for providing us the most updated literature about SDSS-III project. 
We warmly thank as well Jose Beltr\'an for useful comments about BAO calculations.
AA acknowledges financial support from NRF (South Africa).  
AdlCD acknowledges financial support from Marie Curie - Beatriu de Pin\'os contract BP-B00195 Generalitat de Catalunya, ACGC University of Cape Town 
and MINECO (Spain) projects numbers FIS2011-23000, FPA2011-27853-C02-01 and Consolider-Ingenio MULTIDARK CSD2009-00064.
PKSD thanks the NRF for financial support and Queen Mary, University of London for support during the final stages of this work.

\end{document}